\documentclass[aps,prd,preprintnumbers,twocolumn,groupedaddress,showpacs,nofootinbib]{revtex4}
\usepackage{graphicx}
\usepackage{latexsym}
\def\beq{\begin{equation}}
\def\eeq{\end{equation}}
\def\bey{\begin{eqnarray}}
\def\eey{\end{eqnarray}}

\def\lsim{\mathrel{\raise.3ex\hbox{$<$\kern-.75em\lower1ex\hbox{$\sim$}}}}
\def\gsim{\mathrel{\raise.3ex\hbox{$>$\kern-.75em\lower1ex\hbox{$\sim$}}}}

\newcommand{\be}{\begin{equation}}
\newcommand{\ee}{\end{equation}}
\newcommand{\tev}{\ensuremath{\mathrm{\,Te\kern -0.1em V}}\xspace}
\newcommand{\gev}{\ensuremath{\mathrm{\,Ge\kern -0.1em V}}\xspace}
\newcommand{\tevt}{\ensuremath{\mathrm{Te\kern -0.1em V}}\xspace}
\newcommand{\kev}{\ensuremath{\mathrm{\,ke\kern -0.1em V}}\xspace}
\newcommand{\mev}{\ensuremath{\mathrm{\,Me\kern -0.1em V}}\xspace}

\begin{document}

\title{The Isotropic Radio Background and Annihilating Dark Matter}

\author{Dan Hooper$^{1,2}$}
\author{Alexander V.~Belikov$^{3}$}
\author{Tesla E.~Jeltema$^4$}
\author{Tim Linden$^4$}
\author{Stefano Profumo$^4$}
\author{Tracy R.~Slatyer$^5$}
\affiliation{$^{1}$Center for Particle Astrophysics, Fermi National Accelerator Laboratory, Batavia, IL 60510, USA}
\affiliation{$^{2}$Department of Astronomy and Astrophysics, University of Chicago, Chicago, IL 60637, USA}
\affiliation{$^{3}$Institut d'Astrophysique de Paris, UMR 7095, CNRS, UPMC Univ. Paris 06, 98 bis boulevard Arago, 75014 Paris, France}
\affiliation{$^{4}$Department of Physics and Santa Cruz Institute for Particle Physics, University of California, 1156 High Street, Santa Cruz, CA 95064, USA}
\affiliation{$^{5}$School of Natural Sciences, Institute for Advanced Study, Princeton, NJ 08540, USA}

\date{\today}

\begin{abstract}

Observations by ARCADE-2 and other telescopes sensitive to low frequency radiation have revealed the presence of an isotropic radio background with a hard spectral index. The intensity of this observed background is found to exceed the flux predicted from astrophysical sources by a factor of approximately 5-6. In this article, we consider the possibility that annihilating dark matter particles provide the primary contribution to the observed isotropic radio background through the emission of synchrotron radiation from electron and positron annihilation products. For reasonable estimates of the magnetic fields present in clusters and galaxies, we find that dark matter could potentially account for the observed radio excess, but only if it annihilates mostly to electrons and/or muons, and only if it possesses a mass in the range of approximately 5-50 GeV. For such models, the annihilation cross section required to normalize the synchrotron signal to the observed excess is $\sigma v \approx (0.4-30)\times 10^{-26}$ cm$^3$/s, similar to the value predicted for a simple thermal relic ($\sigma v \approx 3\times 10^{-26}$ cm$^3$/s). We find that in any scenario in which dark matter annihilations are responsible for the observed excess radio emission, a significant fraction of the isotropic gamma ray background observed by Fermi must result from dark matter as well. 



\end{abstract}

\pacs{95.85.Bh, 95.85.Fm, 95.35.+d; FERMILAB-PUB-12-072-A}

\maketitle

\section{Introduction}

Over the past several years, evidence indicating the existence of a significant isotropic radio background has been uncovered by a number of instruments.  In 2009, the ARCADE 2 (Absolute Radiometer for Cosmology, Astrophysics and Diffuse Emission) collaboration reported measurements of the absolute sky temperature at a number of frequencies between 3 and 90 GHz~\cite{arcade}. While these measurements are dominated by the cosmic microwave background (CMB) at frequencies above several GHz, they reveal the presence of significant excess power at the lowest measured frequencies~\cite{arcadeinterpretation}. This conclusion is further supported by a number of observations at lower frequencies, as reported by Roger {\it et al.} (0.022 GHz)~\cite{roger}, Maeda {\it et al.} (0.045 GHz)~\cite{maeda}, Haslam {\it et al.} (0.408 GHz)~\cite{haslam}, and Reich and Reich (1.42 GHz)~\cite{reich}. The emission observed by each of these groups is in significant excess of what can be attributed to Galactic emission, or to unresolved members of known extragalactic radio source populations~\cite{arcade,arcadeinterpretation}.


\begin{figure*}[t]
\centering
\includegraphics[angle=0.0,width=6.9in]{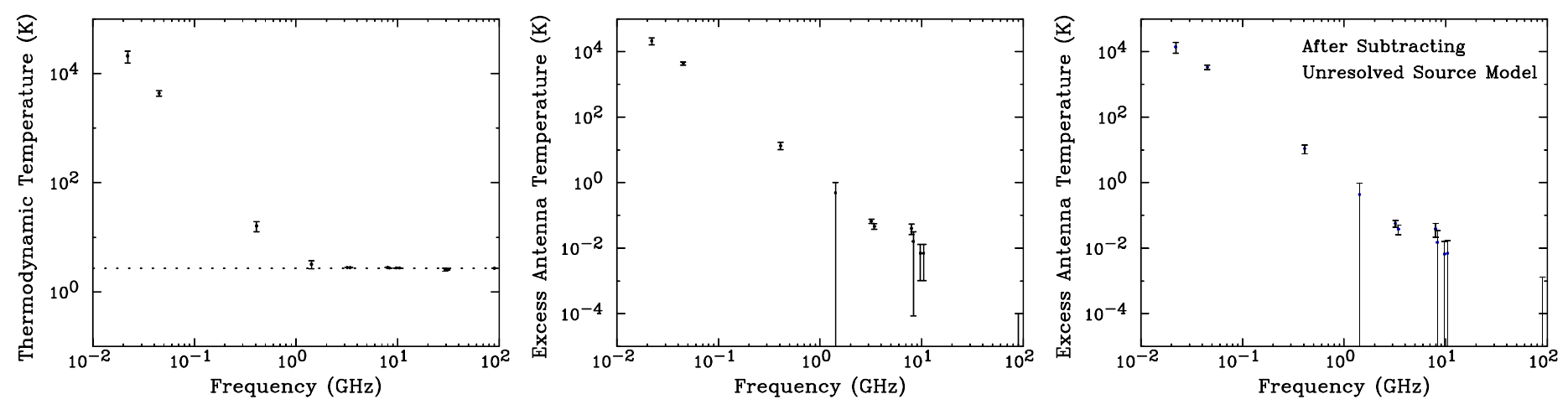}
\caption{Measurements of the isotropic radio background, as reported by Roger {\it et al.} (0.022 GHz)~\cite{roger}, Maeda {\it et al.} (0.045 GHz)~\cite{maeda}, Haslam {\it et al.} (0.408 GHz)~\cite{haslam}, Reich and Reich (1.42 GHz)~\cite{reich}, and by the ARCADE 2 collaboration (3.20 to 90 GHz)~\cite{arcade} (see also, Ref.~\cite{arcadeinterpretation}). In the left frame, the total observed background is shown, including the contribution of the CMB (shown as a dotted line). In the middle frame, the CMB has been subtracted, revealing a power-law spectrum of isotropic emission. In the right frame, an estimate of the contribution from unresolved radio sources has been subtracted as well. See text for more details.}
\label{temp}
\end{figure*}

In the left frame of Fig.~\ref{temp}, we plot the measured spectrum of the isotropic radio background (after removing galactic foregrounds and resolved radio sources)~\cite{arcadeinterpretation}. At frequencies below a few GHz, these measurements significantly exceed the uniform temperature of the CMB (which is shown as a horizontal dotted line). In the center frame, this spectrum is shown after subtracting the contribution of the CMB, revealing a power-law like spectrum with an index of approximately $-2.6$ ($-1.6$ in $dN_{\gamma}/dE_{\gamma}$ units) extending over at least three orders of magnitude in frequency, between $\sim$20 MHz and $\sim$10 GHz~\cite{arcade}. And while a fraction of this emission is expected to originate from faint and currently unresolved radio sources, estimates of this contribution based on deep surveys from the Very Large Array (VLA), among other observations, find that such sources should account for only approximately 20\% of the observed isotropic radio emission.  In the right frame of Fig.~\ref{temp}, we show the spectrum after subtracting the estimated contribution from unresolved radio sources, as described in Ref.~\cite{arcadeinterpretation}.

The origin of the excess isotropic radio emission is currently unknown. Various systematic effects to explain this excess have been ruled out~\cite{arcade}, and galactic origins (such as free-free or galactic synchrotron emission) are difficult to accommodate~\cite{galactic,sources,arcadeinterpretation}. Standard astrophysical sources, such as radio quiet quasars, radio supernovae, and diffuse emission from clusters, have been considered but appear to be unable to account for the excess emission~\cite{sources,Vernstrom:2011xt}. 

Any astrophysical objects capable of producing the observed isotropic excess without exceeding the observed number of resolved radio sources must be both highly numerous and very faint~\cite{Gervasi:2008rr,Vernstrom:2011xt}. The best known example of such a source class are star-forming galaxies. However, in order for star-forming galaxies to account for the entire isotropic radio flux without exceeding the amount of emission observed at infrared (IR) wavelengths, the ratio of radio-to-IR emission must be increased by approximately a factor of 5 at high redshifts over what is observed from local sources. Such strong evolution seems in conflict with current measurements~\cite{Ponente:2011se}. Furthermore, measurements of the isotropic gamma ray flux also strongly constrain such a scenario~\cite{Lacki:2010uz}.

An alternative possibility for the origin of the excess isotropic radio emission is annihilating dark matter~\cite{Fornengo:2011cn,Fornengo:2011xk}. Electrons and positrons produced in dark matter annihilations lose energy through both synchrotron and inverse Compton processes, the former of which contributes to the radio background. Radio synchrotron emission has long been considered as a probe of dark matter annihilations in a number of contexts: from the Inner Milky Way~\cite{haze}, the Galactic Center~\cite{gcradio}, the Milky Way's non-thermal radio filaments~\cite{filaments}, galaxy clusters~\cite{clusterradio}, and nearby dwarf galaxies and other substructures~\cite{dwarfradio}. Dark matter halos are predicted to be very numerous and faint radio sources -- every dark matter halo large enough to contain a significant magnetic field constitutes such a source -- making them ideal candidates to generate the observed radio background~\cite{Fornengo:2011cn}. And while individual extragalactic halos are not expected to be bright enough to be resolved as radio sources, the sum of all halos could collectively constitute a bright source of isotropic radio emission. Furthermore, unlike most astrophysical sources of radio emission, some models of dark matter annihilation produce comparatively little emission at IR or gamma-ray wavelengths, and thus are relatively unconstrained by such observations (especially in the case of annihilations to $e^+ e^-$ or $\mu^+ \mu^-$).

In this paper, we calculate the spectrum of radio emission from dark matter annihilations taking place within the halos of galaxies and galaxy clusters, and determine in what cases such annihilations could potentially account for the observed isotropic radio excess. We find that among a relatively narrow range of dark matter models, it may be possible to explain the observed emission. In particular, to produce the observed radio spectrum while also evading constraints from gamma-ray observations (of both the isotropic gamma ray background and local dwarf galaxies), we find that we must consider models in which the dark matter particles are somewhat light ($m_{\rm DM} \sim 4-50$ GeV) and annihilate significantly to either $e^+ e^-$ or $\mu^+ \mu^-$ final states. In such models, the overall normalization of the radio excess requires dark matter annihilation cross sections which are typically similar to the value predicted for a simple thermal relic ($\sigma v \sim 3 \times 10^{-26}$ cm$^3$/s). In all of the dark matter models we found to be capable of producing the observed radio excess, dark matter annihilation products are also predicted to make up a significant fraction of the isotropic gamma ray background.

\section{The Contribution From Dark Matter Annihilations to the Extragalactic Radio Background}
\label{calc}

In most models of weakly interacting massive particles (WIMPs), dark matter annihilations deposit energy into a combination of Standard Model particles, including gamma rays, neutrinos, protons, electrons, and their antiparticles. Energetic electrons and positrons produced through this process subsequently lose energy through a combination of synchrotron and inverse Compton scattering.  While inverse Compton scattering leads to the production of gamma rays and X-rays~\cite{Profumo:2009uf,Belikov:2009cx}, synchrotron processes with micro-Gauss-scale magnetic fields yield photons at radio and microwave wavelengths. 

The dark matter annihilation rate per volume, at redshift $z$, is given by:
\begin{equation}
R(z)=\int  \frac{dn}{dM}(M,z) (1+z)^3 dM \frac{\sigma v}{2 m^2_{\rm DM}} \int \rho^2(\vec{x},M,z) dV,
\end{equation}
where $dn/dM$ is the differential comoving number density of dark matter halos of mass $M$, $\sigma v$ is the dark matter annihilation cross section (multiplied by the relative velocity of the two WIMPs), $m_{\rm DM}$ is the mass of the dark matter particle, and $\rho(\vec{x},M,z)$ is the density of dark matter at a location $\vec{x}$ within a halo of mass $M$.

We calculate the number density of dark matter halos of a given mass as a function of redshift, $\frac{dn}{dM}(M,z)$ as described in Ref.~\cite{Belikov:2009cx}, using standard values for cosmological parameters. To describe the distribution of dark matter within a given halo, we adopt a generalized Navarro-Frenk-White (NFW) form~\cite{nfw} ($\gamma=1$ being the canonical NFW value):
\begin{equation}
\rho(r) \propto \frac{1}{(r/R_s)^{\gamma}(1+r/R_s)^{3-\gamma}},
\end{equation}
where $R_s$ is the scale radius of the halo, which is related to the virial radius in terms of the halo concentration, $c\equiv R_{\rm vir}/R_s$. For default parameters, we adopt $\gamma=1.0$ for the largest halos ($10^{13}-10^{15} M_{\odot}$) and $\gamma=1.3$ for halos with masses in the range of $10^{10}-10^{13} M_{\odot}$ (as favored by hydrodynamical simulations which account for the effects of baryonic contraction~\cite{Gnedin:2011uj} and as observed among early-type galaxies~\cite{Chae:2012qq}). To determine the concentration of a halo of a given mass and at a given redshift, $c(z,M)$, we use the model of Bullock {\it et al.}~\cite{bullock} (alternatively, using the mass-concentration relationship of Ref.~\cite{MunozCuartas:2010ig} modifies our results only slightly). We also account for halo-to-halo variations in concentration, which are taken to follow a log-normal distribution with $\Delta(\log_{10}c)=0.2$~\cite{halotohalo}.


Within a given dark matter halo, there are a large number of smaller subhalos, each of which contribute to the total annihilation rate. In our calculations, we adopt two treatments to describe the effect of substructure on the annihilation rate. First we use a generalization of the analytic model of Kamionkowski {\it et al.}~\cite{Kamionkowski:2010mi}, which is calibrated to the results of the Via Lactea simulation~\cite{vialactea}. In particular, we adopt an annihilation rate boost factor which follows the radial profile given by their Fig.~4, and provides an overall integrated boost of 17 for a Milky Way-sized halo~\cite{Kamionkowski:2010mi}. For larger and smaller halos, we scale the overall boost factor by $M^{0.39}_{\rm halo}$~\cite{boostscaling} and adopt a radial dependence which follows Ref.~\cite{Kamionkowski:2010mi}, scaled to the virial radius of the halo. As a second case, we adopt a similar substructure model, but normalized to the results of the Aquarius simulation~\cite{aquarius,boostscaling}. This substructure model yields an overall boost factor that is approximately 4.5 larger than that of the Via Lactea model. Note that these substructure models rely on extrapolations over many orders of magnitude in halo mass (from the current resolution of simulations to the smallest halo masses, taken to be $10^{-6} \, M_{\odot}$), and are thus quite uncertain. On the other hand, because most of the annihilation boost from substructure occurs in the outer regions of halos where the magnetic fields are fairly weak, the flux of synchrotron emission is only modestly impacted by substructure, in particular when compared to the impact of substructure on gamma ray emission (neglecting substructure entirely reduces the radio flux only by a factor of a few, relative to the Via Lactea case).

The magnetic fields found in galaxies and and galaxy clusters are produced by complicated and not easily modelled baryonic physics and are difficult to reliably estimate. They are also rather poorly measured and are thus only modestly constrained by observations. In our calculations, we adopt a simple and reasonable estimate for the distributions of magnetic fields, while acknowledging the considerable uncertainties being introduced. We will return to this issue in Sec.~\ref{uncertainties}. For the largest halos ($10^{13}-10^{15} M_{\odot}$), corresponding to the scales of galaxy clusters, we adopt the following parameterization~\cite{clusterB}:
\begin{equation}
B(r) = B_{0} \bigg[\,1+\bigg(\frac{r}{r_c}\bigg)^2\,\bigg]^{-3 \beta \eta/2},
\label{bcluster}
\end{equation}
where $3\beta$ (which we take to be $3 \times 0.6$) is the slope of the cluster gas profile at $r \gg r_c$ and $r_c$ is the core radius of the gas distribution (taken to be 5\% of the halo's virial radius, $r_c=0.05 \, R_{\rm vir}$). We adopt a central field strength of $B_0=10$ $\mu$G which falls off according to $\eta=0.5$.

For magnetic fields in somewhat smaller, galaxy-scale halos ($10^{10}-10^{13} M_{\odot}$), we start with a double exponential model:
\begin{eqnarray}
B(r,z)=B_0 \exp(-R/R_0) \exp(-z/z_0),
\label{bgalaxy}
\end{eqnarray}
with $B_0=7.6 \mu$G, $R_0=0.17 \, R_{\rm vir}$, and $z_0=0.01 \, R_{\rm vir}$. In light of observations which indicate the presence of rather strong magnetic fields in the inner region of the Milky Way~\cite{highB}, we replace the above parameters within the inner several kiloparsecs of galaxy-scale halos by $B_0=35$ $\mu$G, $R_0=0.017 \, R_{\rm vir}$, and $z_0=0.006 \, R_{\rm vir}$ (at a given location, we adopt the parameter set which yields the largest value of $B$, allowing for a continuous transition between the two sets of parameters).

Although we neglect any isolated (non-substructure) halos smaller than $10^{10} M_{\odot}$ in our calculations, we note that if we had included halos in the mass range of $10^{10} M_{\odot} > M > 10^{8} M_{\odot}$ with 1 $\mu$G fields, they would contribute only on the order of 1\% of the total radio emission.

\begin{figure*}[t]
\centering
\includegraphics[angle=0.0,width=5.4in]{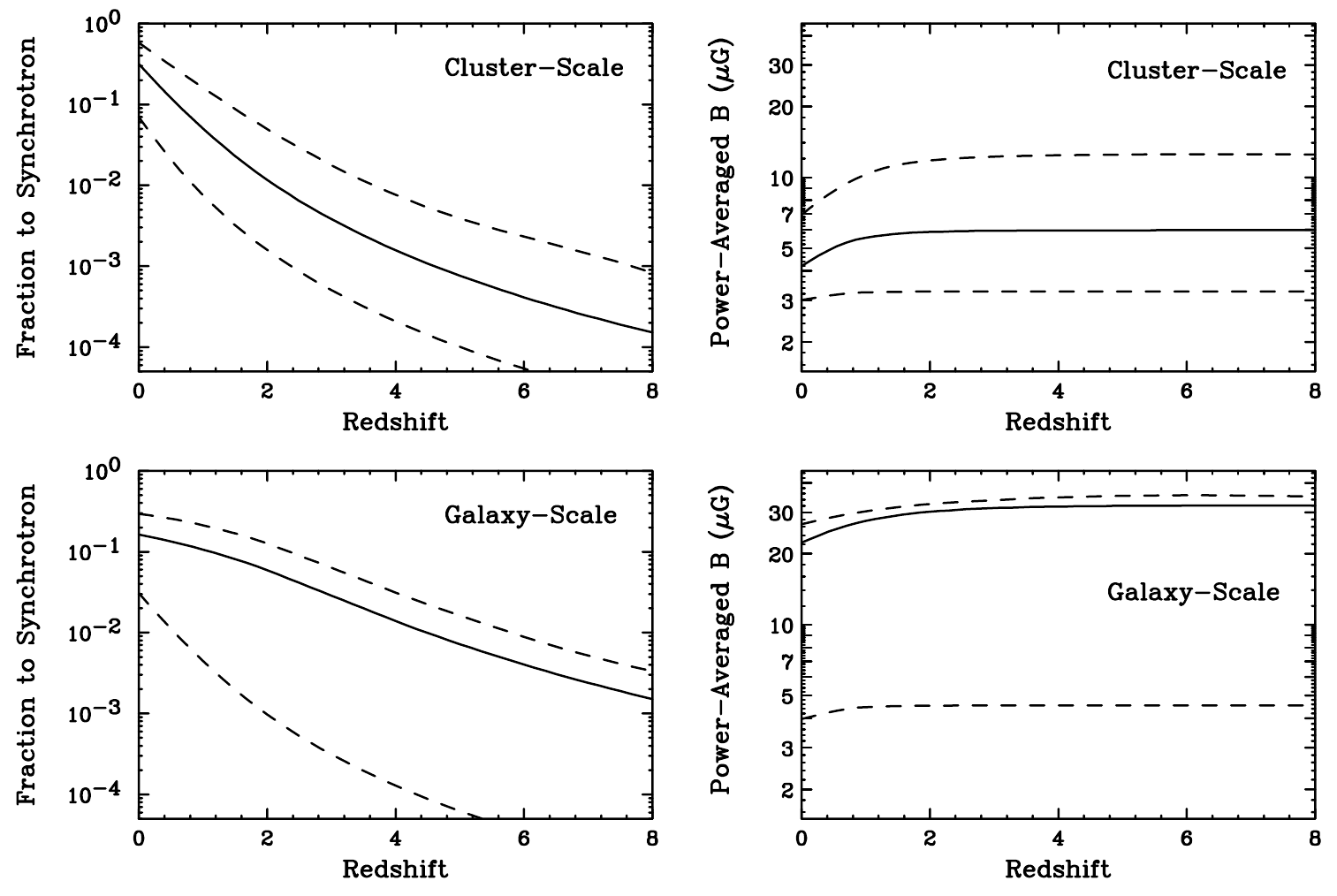}
\caption{The fraction of energy in electron/positron dark matter annihilation products which is lost to synchrotron emission (left) and the power-averaged magnetic field strength which contributes to that synchrotron emission (right). In the upper and lower frames, we show result for our models for the largest dark matter halos ($M>10^{13}\, M_{\odot}$) and somewhat smaller halos ($10^{10}-10^{13}\, M_{\odot}$), respectively. In each frame, the solid lines denote the results for our default magnetic field and dark matter distribution models. The dashed lines represent the range of values found using the variety of models described in Sec.~\ref{uncertainties}. In each case shown, we use the Via Lactea substructure model.}
\label{fracB}
\end{figure*}

In calculating the contribution of dark matter annihilation products to the isotropic radio background, one might worry that magnetic fields in dark matter halos may evolve with redshift, potentially contributing less from more distant, and thus older, halos. As it turns out, however, relatively nearby halos overwhelmingly dominate this signal (halos at $z< 1$ provide approximately 98\% of the total contribution to the radio background in most of the models considered here). The primary reason for this can be seen in the left frames of Fig.~\ref{fracB}, where we plot the fraction of energy in electron/positron dark matter annihilation products which is emitted as synchrotron (as opposed to inverse Compton scattering with the cosmic microwave background), as a function of redshift. Due to the rapid evolution of the CMB energy density with redshift, $\rho_{\rm CMB}\propto (1+z)^4$, synchrotron losses play only a minor role for energetic electrons in cosmologically distant halos. Furthermore, evidence also suggests that magnetic fields in high redshift galaxies and galaxy clusters are not significantly lower than those found in local halos~\cite{Arshakian:2009dr}.

The diffuse and isotropic spectrum of synchrotron emission from dark matter annihilations taking place throughout the universe is described by:
\begin{equation}
\frac{d\phi_{\rm syn}}{dE_{\rm syn}} = \frac{\sigma v}{8 \pi}\frac{c}{H_0} \frac{\bar{\rho}^2_{\rm DM}}{m^2_{\rm DM}} \int dz (1+z)^3 \frac{\Delta^2(z)}{h(z)}\frac{dN_{\rm syn}}{dE_{\rm syn}}(E_{{\rm syn},0}(1+z)),
\end{equation}
where $\bar{\rho}_{\rm DM}$ is the average cosmological density of dark matter, $h(z)=\sqrt{\Omega_{\Lambda}+\Omega_{M}(1+z)^3}$, and $\Delta^2(z)$ is the average squared overdensity of dark matter, which is determined from a combination of the halo mass function, halo profiles, and contributions from substructure. $dN_{\rm syn}/dE_{\rm syn}$ is the spectrum of synchrotron emission per dark matter annihilation, which depends on the spectrum of electrons/positrons injected, and on the magnetic field strength and other characteristics of the environment in which the annihilations take place. The steady-state distribution of cosmic ray electrons and positrons can be approximately determined by solving the diffusion-loss equation:
\begin{eqnarray}
0&=&\vec{\nabla}\cdot [K(\vec{x},E_e) \vec{\nabla} \frac{dN_e}{dE_e}(\vec{x},E_e)]\nonumber \\ 
&+&\frac{\partial}{\partial E_e} [b(\vec{x},E_e) \frac{dN_e}{dE_e}(\vec{x},E_e)]+Q(\vec{x},E_e),
\end{eqnarray}
where $K(\vec{x},E_e)$ is the diffusion constant and $b(\vec{x},E_e)$ is the energy loss rate. In the energy range we are most interested in, energy losses are dominated by inverse Compton and synchrotron processes:
\begin{eqnarray}
b(\vec{x},E_e) &=& \frac{4}{3}\sigma_T \rho_{\rm rad} \bigg(\frac{E_e}{m_e}\bigg)^2 + \frac{4}{3}\sigma_T \rho_{\rm mag} \bigg(\frac{E_e}{m_e}\bigg)^2   \\
\approx 1.02 &\times& 10^{-16} \, {\rm GeV/s} \, \times \bigg(\frac{E_e}{{\rm GeV}}\bigg)^2 \nonumber \\
&\times& \bigg[\bigg(\frac{\rho_{\rm rad}}{{\rm eV/cm}^3}\bigg)+\bigg(\frac{B}{6.35\, \mu G}\bigg)^2        \bigg], \nonumber
\end{eqnarray}
where $\rho_{\rm rad}$ is the energy density of radiation and $B$ is the magnetic field strength. Note that in magnetic fields stronger than $3.25 \, \mu G$, energy emitted as synchrotron will exceed inverse Compton losses to the cosmic microwave background (at $z=0$). 

In large halos, such as those of galaxy clusters, diffusion can safely be neglected (variations in the magnetic field strength and dark matter density occur on length scales much larger than electrons and positrons travel before losing their energy). In this case, the steady-state equilibrium number density of electrons/positrons at a given location is given by:
\begin{equation}
\frac{dN_e}{dE_e}(E_e) = \frac{\sigma v \rho_{\rm DM}^2}{2 m^2_{\rm DM} b(E_e)} \int^{\infty}_{E_e} dE'_e \frac{dN_{e,{\rm Inj}}}{dE}(E'_e),
\end{equation}
where $b(E_e)$ again is the energy loss rate and $dN_{e, {\rm Inj}}/dE$ is the spectrum of electrons/positrons injected per dark matter annihilation. In smaller halos, electrons and positrons can potentially diffuse into regions with non-negligibly differing magnetic fields before losing most of their energy, which can alter the resulting synchrotron spectrum, and the fractions of energy lost to synchrotron and Inverse Compton. To estimate the scale of this effect, we calculate the typical distance that an electron diffuses before losing the majority of its energy to synchrotron. In a magnetic field of strength $B$, an electron of energy $E_e$ will lose half of its energy in a time equal to $t_{\rm loss}=E_e/b(E_e)\approx 4 \times 10^{14} \, {\rm s} \times (10\,{\rm GeV}/E_e)(10 \, \mu{\rm G}/B)^2$. During that time, a typical electron will diffuse a distance of $L_{\rm Dif} \approx \sqrt{K(E_e) t_{\rm loss}(E_e)}$, which for a diffusion coefficient of $K(E_e) \approx 5\times 10^{28} \, {\rm cm}^2/{\rm s}$ yields $L_{\rm Dif} \approx 1.45 \, {\rm kpc} \times (10 \,{\rm GeV}/E_e)^{0.5} (10 \, \mu{\rm G}/B)$. As such a distance scale is not entirely irrelevant in Galaxy-scale halos (but is irrelevant on cluster-scales), we will attempt to more quantitatively assess the impact of diffusion on the overall synchrotron spectrum and total power that results from dark matter annihilations in Galaxy-scale halos. To accomplish this, we employed a Monte Carlo technique to propagate electrons and positrons generated in dark matter annihilation over the region through which they diffuse in one energy-loss time (calculated using the combined magnetic field and CMB energy densities at the location at which the annihilation takes place), and then obtain the average ratio of the magnetic field to CMB energy densities in this region. Employing a diffusion constant of 5~x~10$^{28}$~cm$^2/$s, using both the magnetic field and dark matter density models described above, we find that 10 GeV electrons generated in dark matter annihilations deposit approximately 35\% less power to synchrotron radiation than is found when diffusion is not considered. For our default parameters, we find that approximately 56\% (86\%) of the overall radio signal comes from halos larger than $10^{13} \, M_{\odot}$ when using the Via Lactea (Aquarius) substructure model. For such large halos, the effects of diffusion are negligible. We thus estimate the overall suppression of the dark matter contribution to the radio background due to diffusion to be on the order of $\sim$5-15\%, which is a small effect when compared to the significant uncertainties in the dark matter distribution and magnetic field models being used. Diffusion will also lead to a slight softening of the resulting synchrotron spectrum, although this effect is small when compared to the uncertainties in the magnetic fields.

From the steady-state spectrum of electrons and positrons, $dN_e/dE_e(E_e)$, we can calculate the spectrum of synchrotron emission per annihilation:
\begin{equation}
\frac{dN_{\rm syn}}{dE_{\rm syn}} = \frac{4\sqrt{3}e^3 B m^2_{\rm DM}}{m_e c^2 \sigma v \rho_{\rm DM}^2} \int^{m_{\rm DM}}_{m_e} dE' F(E/E_c)  \frac{dN_e}{dE_e}(E').
\end{equation}
Note that the quantity $dN_e/dE_e$ is proportional to the annihilation cross section and the square of the number density of dark matter particles, which cancel with those quantities in the first portion of this expression.  The spectral shape of synchrotron emission is described by the function $F(x)\equiv x \int^{\infty}_x dz K_{5/3}(z)$, where $K_{5/3}$ is the modified Bessel function of the second kind. At high and low energies/frequencies, the limiting behavior of $F(x)$ is given by:
\begin{eqnarray}
F(x)\approx \frac{4\pi}{\sqrt{3} \,\, \Gamma(1/3)} \bigg(\frac{x}{2}\bigg)^{1/3}, \,\,\,\,\,\,\,  x \ll 1 
\end{eqnarray}
and
\begin{eqnarray}
F(x)\approx \bigg(\frac{\pi}{2}\bigg)^{1/2} x^{1/2}\, \exp(-x), \,\,\,\,\,\,\,  x \gg 1.
\end{eqnarray}
We know of no simple closed form to describe the behavior of $F(x)$ at intermediate energies. The critical frequency, $\nu_c$ (and thus the critical energy, $E_c=h\nu_c$), of synchrotron emission is determined by the electron energy and by the strength of the magnetic field, $\nu_c= B E^2_e \times (3ce/4\pi m^3_e c^6)$.

With this formalism, we turn in the following section to the resulting spectrum and flux of the isotropic radio background generated from annihilating dark matter particles.

\begin{figure*}[t]
\centering
\includegraphics[angle=0.0,width=3.05in]{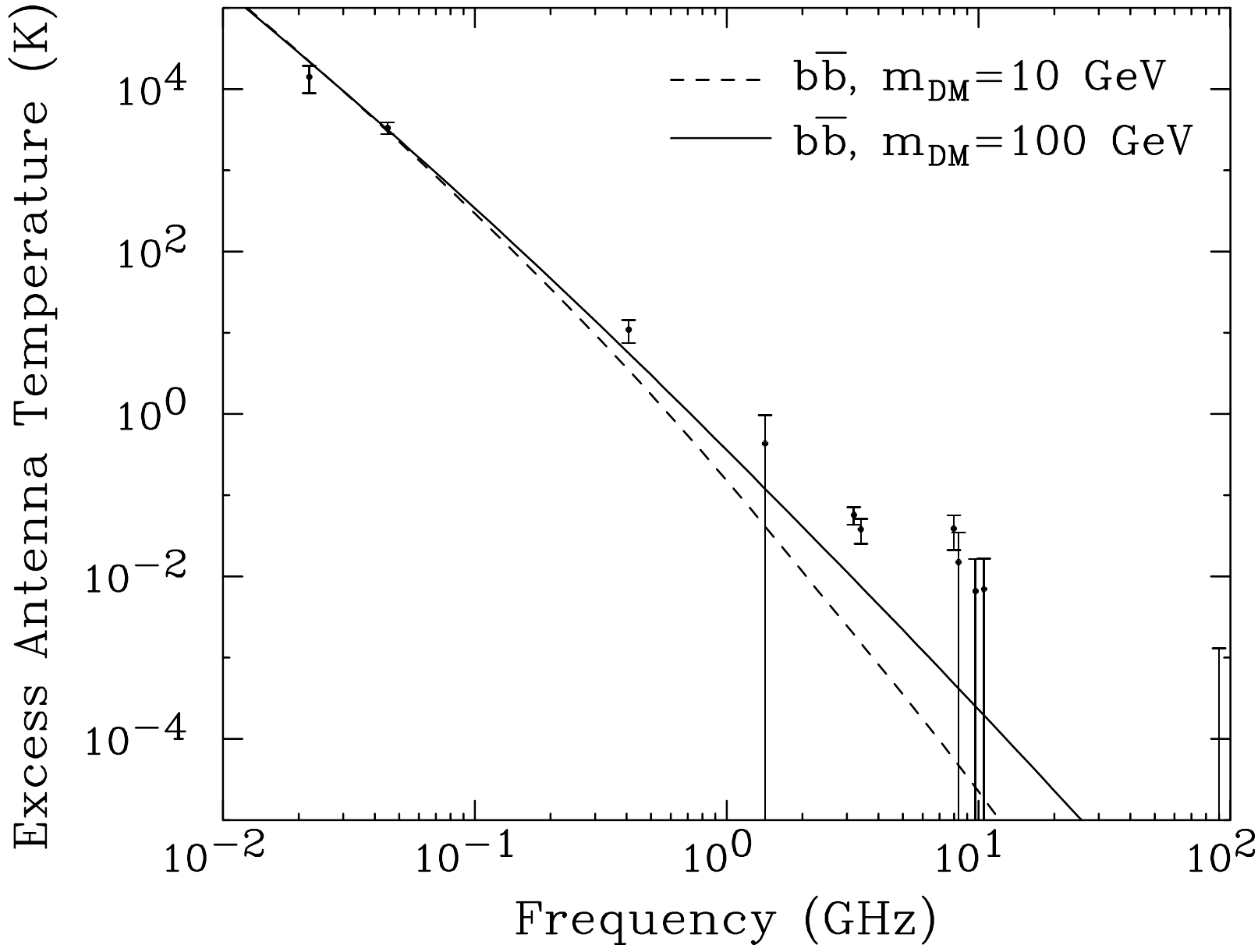}
\hspace{0.3cm}
\includegraphics[angle=0.0,width=3.05in]{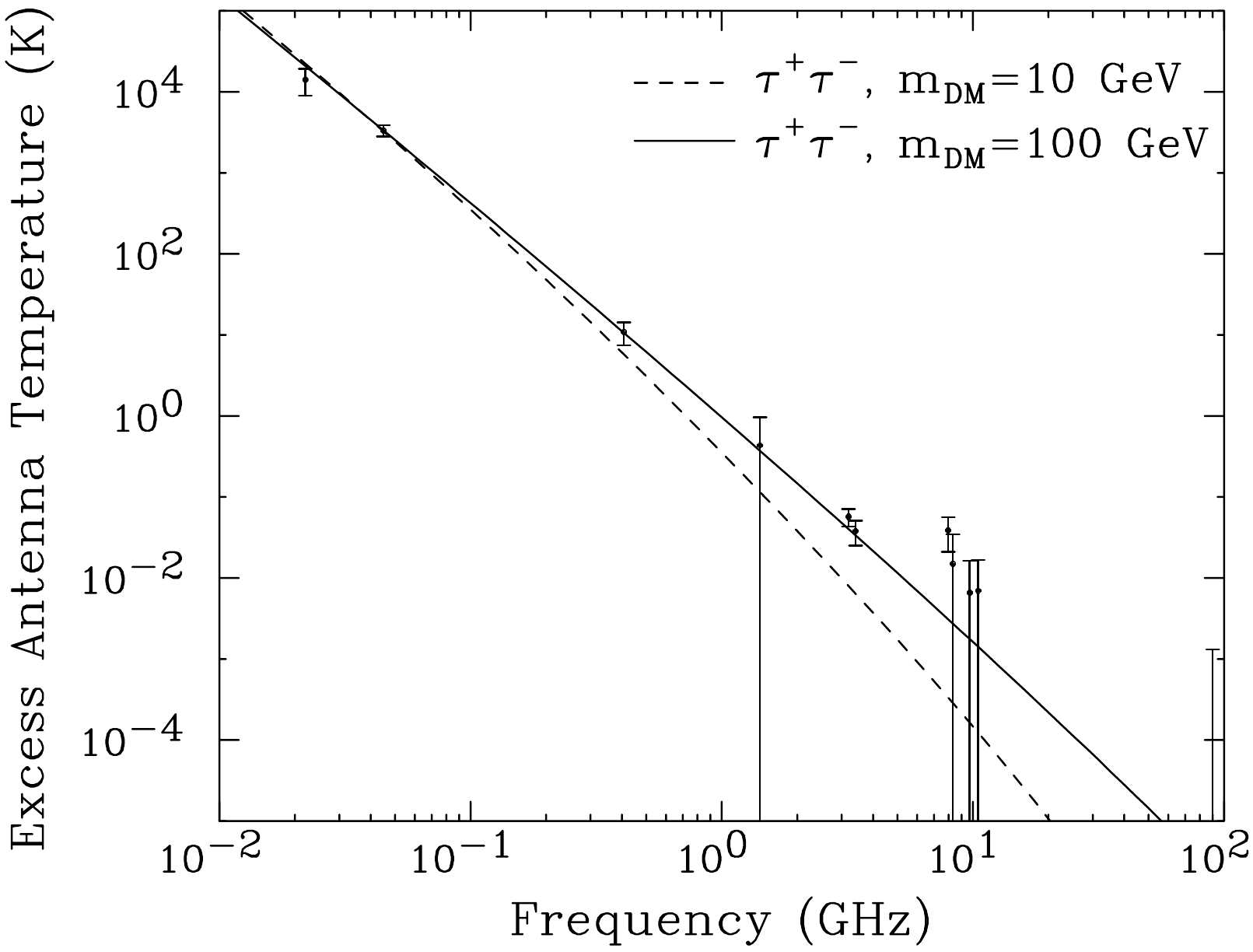}\\
\includegraphics[angle=0.0,width=3.05in]{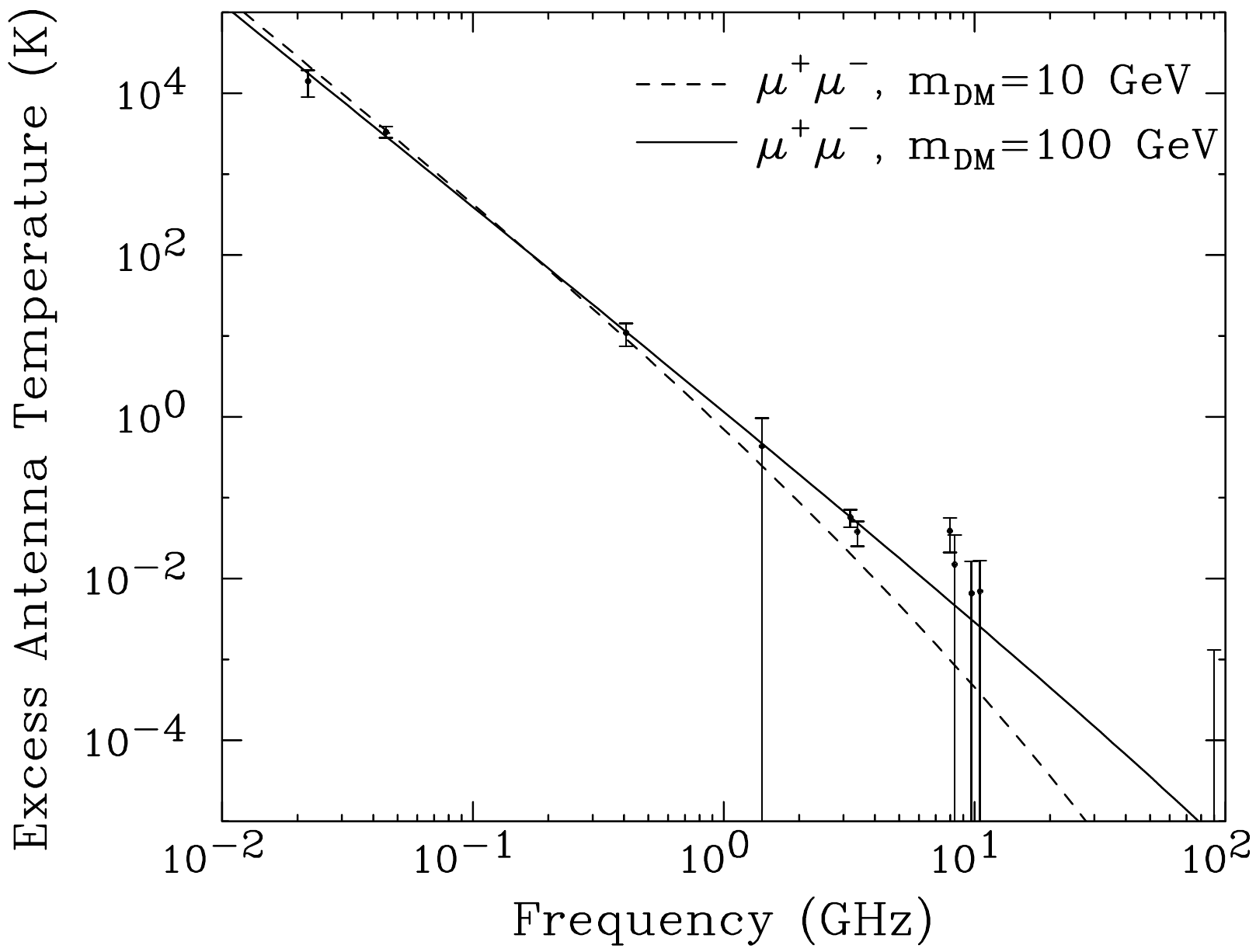}
\hspace{0.3cm}
\includegraphics[angle=0.0,width=3.05in]{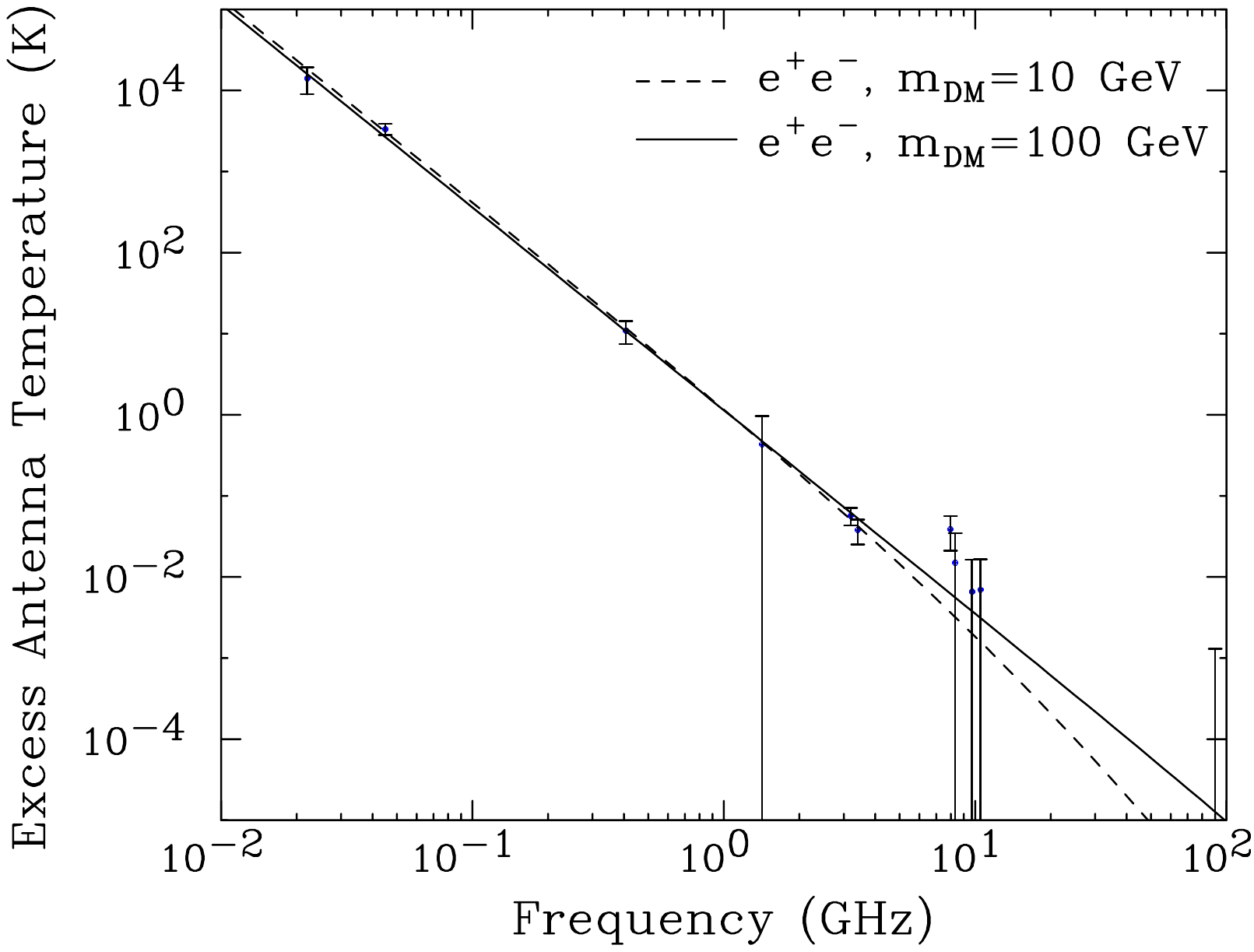}
\caption{The contribution to the isotropic radio background from annihilating dark matter particles with masses of 10 or 100 GeV, and annihilating to $b\bar{b}$, $\tau^+\tau^-$, $\mu^+ \mu^-$, or $e^+ e^-$. In each case, we have used our default magnetic field models and dark matter distributions, and substructure as favored by the Via Lactea simulation (see Sec.~\ref{calc}). In each case, the annihilation cross section was selected to best fit the observed radio excess. In the case of annihilations to $b\bar{b}$, we find a very poor fit to the observed spectrum, corresponding to $\chi^2=44.2$ and 34.7 for 10 and 100 GeV masses, respectively, over 12 degrees of freedom. We find similarly poor fits for annihilations to other hadronic final states, and for annihilations to gauge or higgs bosons. Other choices of the magnetic field and substructure models do not significantly alter this conclusion. In contrast, annihilations to leptonic final states can provide a good fit to the spectrum of the observed excess. Annihilations of a 10 GeV dark matter particle to taus, muons, or electrons yield a $\chi^2$ of 36.0, 24.8, and 12.3, respectively. For a 100 GeV mass, these channels yield a $\chi^2$ of 14.4, 12.1, and 13.1. The quality of these fits can change somewhat with variations of the magnetic field model, although realistic models universally favor leptonic annihilation channels.}
\label{spec}
\end{figure*}

\begin{figure*}[t]
\centering
\includegraphics[angle=0.0,width=3.05in]{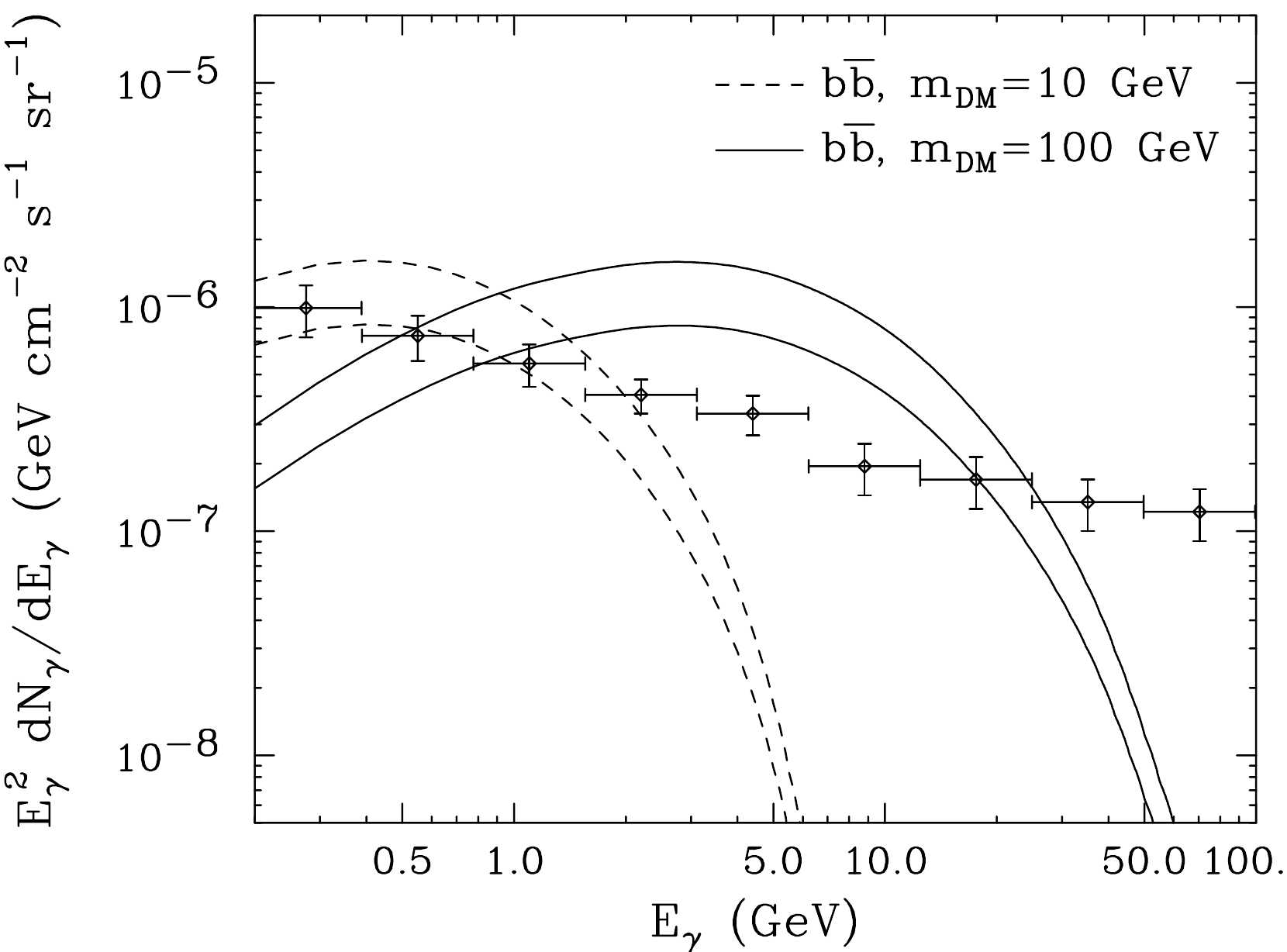}
\hspace{0.3cm}
\includegraphics[angle=0.0,width=3.05in]{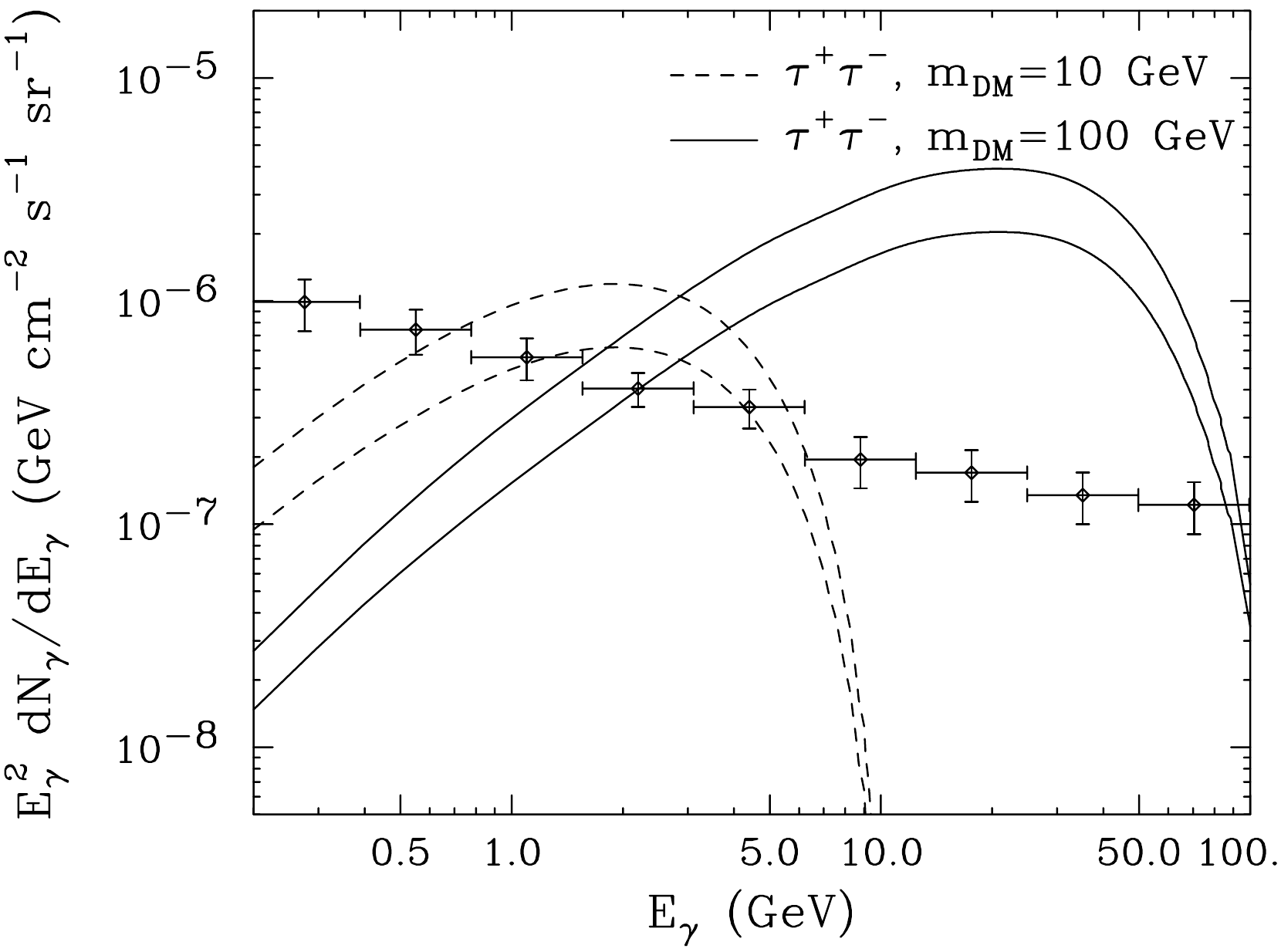}\\
\includegraphics[angle=0.0,width=3.05in]{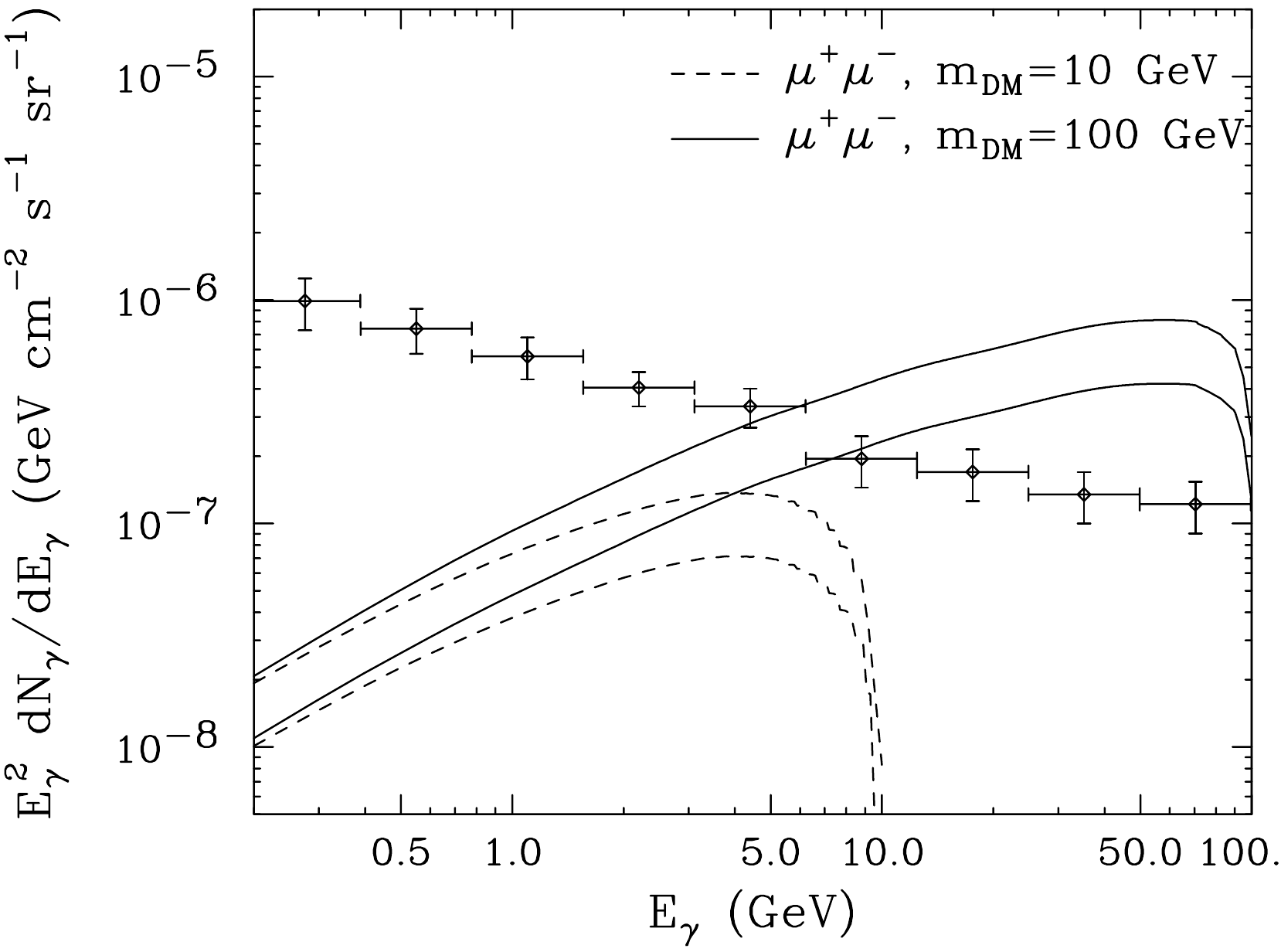}
\hspace{0.3cm}
\includegraphics[angle=0.0,width=3.05in]{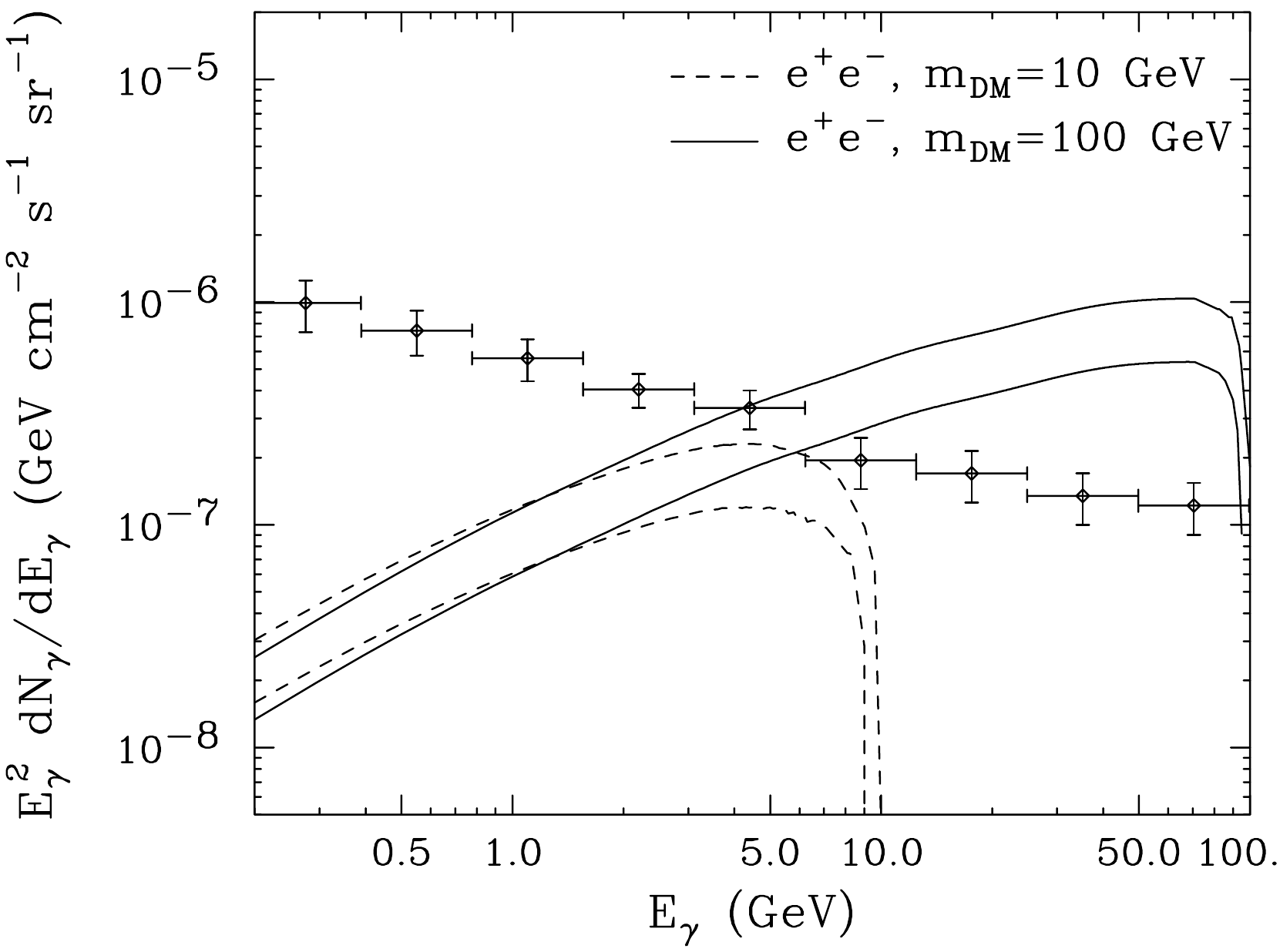}
\caption{The contribution to the isotropic gamma-ray background from annihilating dark matter, in the same models as shown in Fig.~\ref{spec}. In each case, the upper curve use an extrapolatation to determine the concentration of halos lighter than $10^5\, M_{\odot}$, whereas the lower curve fixes the concentration below this mass. When these contributions are compared to the isotropic gamma-ray background observed by Fermi, we see that annihilations of heavy ($m_{\rm DM} \gg 10$ GeV) dark matter particles, or dark matter particles which annihilate primarily to hadronic or $\tau^+ \tau^-$ channels, are unable to produce the observed radio background without also exceeding the observed gamma ray spectrum. }
\label{gamma}
\end{figure*}

\begin{figure*}[t]
\centering
\includegraphics[angle=0.0,width=3.55in]{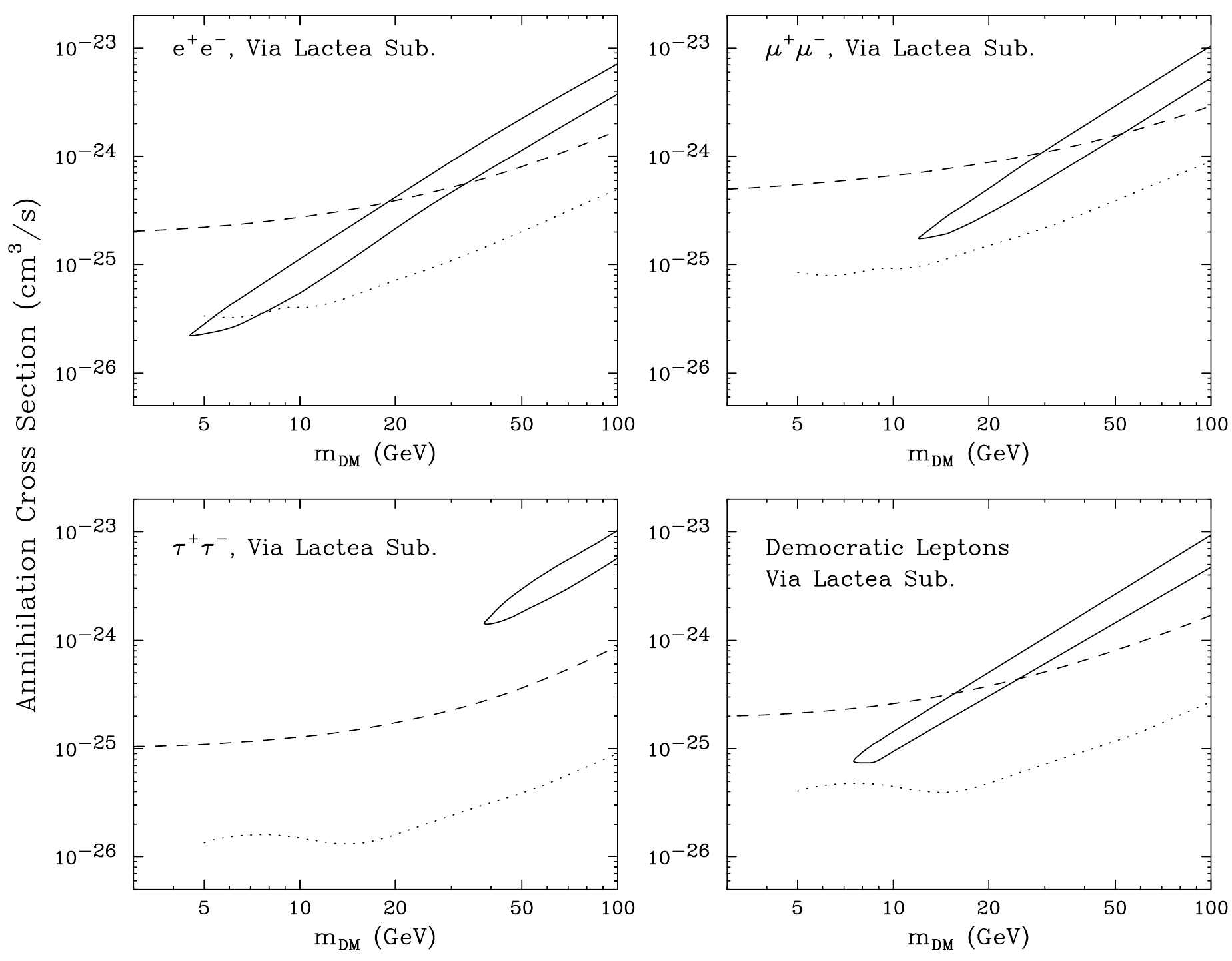}
\hspace{-0.5cm}
\includegraphics[angle=0.0,width=3.55in]{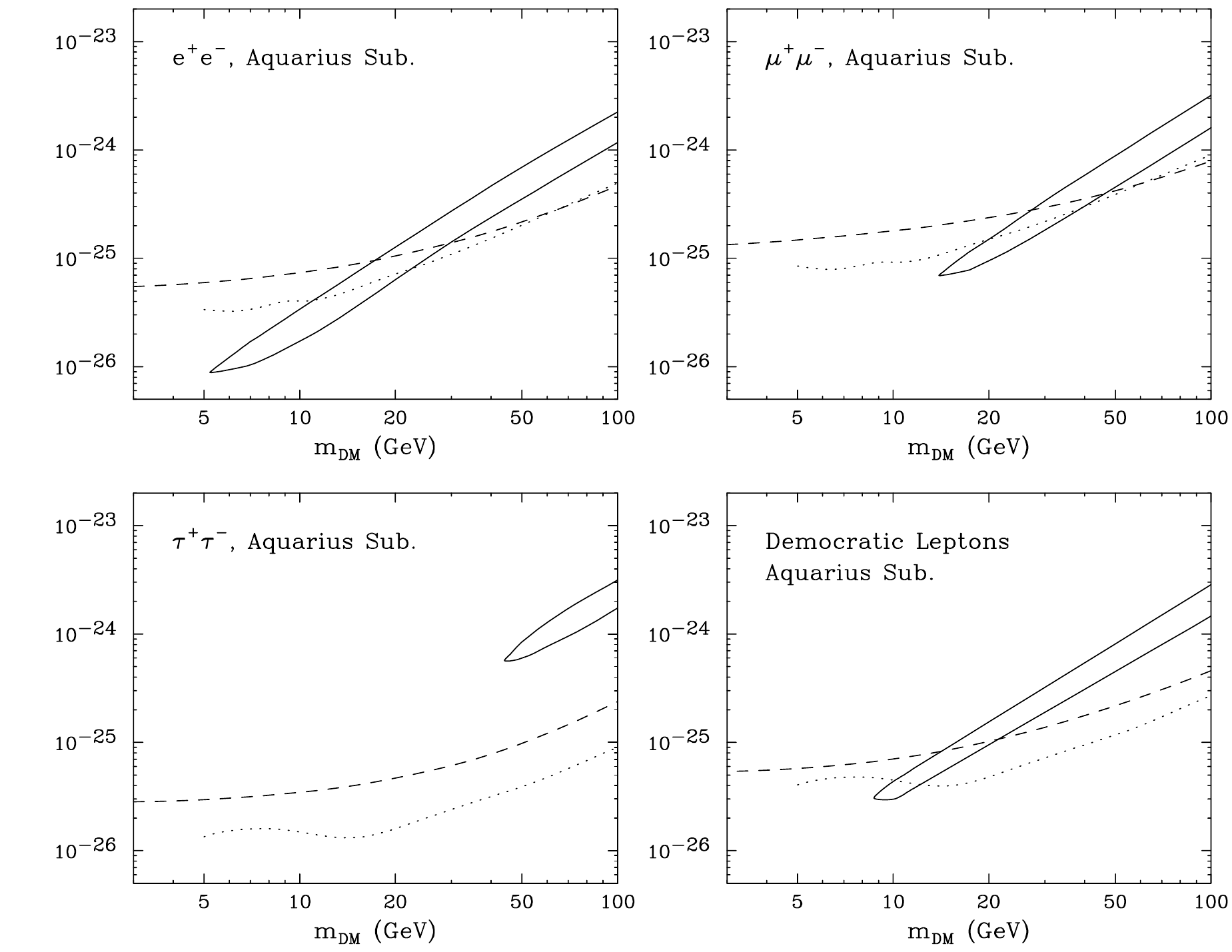}
\caption{The range of dark matter masses and annihilation cross sections which can provide a good fit to the spectrum of the isotropic radio excess, using our default magnetic field and dark matter distribution models (solid line, 95\% confidence regions). No models annihilating to quarks or gauge bosons were found to provide a good fit to the observed spectrum. Also shown are gamma ray constraints derived from the isotropic gamma ray background (dashed lines) and dwarf spheroidal galaxies (dotted lines). See text for more details.}
\label{fits}
\end{figure*}

\section{Results}
\label{results}

In Fig.~\ref{spec}, we plot the spectral contribution of dark matter to the isotropic radio background for a number of dominant annihilation channels and dark matter masses. In each case, we have chosen the annihilation cross section to fit the overall normalization of the observed excess isotropic background, and have adopted the Via Lactea-motivated substructure model (see Sec.~\ref{calc}). In the upper left frame, we see that dark matter annihilations to $b\bar{b}$ predict a spectrum of synchrotron emission that is too soft to account for the observed radio background (we find fits of $\chi^2=44.2$ and 34.7 for $m_{\rm DM}=$10 and 100 GeV, respectively, over 12 degrees of freedom). We find similarly poor fits for annihilations to other hadronic final states, or to gauge or higgs bosons. Although stronger magnetic fields could potentially harden this spectrum and improve this fit to some extent, no plausible magnetic field models can bring this case into concordance with the observed radio spectrum.

We are therefore forced to conclude that in order for dark matter to account for the observed radio spectrum, the annihilations must proceed largely to leptons, as shown in the other three frames of Fig.~\ref{spec}. In these cases, it is much easier to accommodate the observed spectral shape. In particular, 10 GeV dark matter particles annihilating to electrons and 100 GeV particles annihilating to taus, muons, or electrons, each provide good fits to the observed spectrum ($\chi^2=12.3$, 14.4, 12.1, and 13.1, respectively). Several of these cases, however, also predict gamma ray fluxes which exceed the isotropic gamma ray background, as reported by the Fermi Collaboration~\cite{fermibg} (see also, Ref.~\cite{lightbg}). In Fig.~\ref{gamma}, we plot the spectrum of isotropic gamma ray emission predicted for the same dark matter models shown in Fig.~\ref{spec}. In each case, two lines are shown: upper curves corresponds to Bullock-like concentrations, extrapolated down to $10^{-6} \, M_{\odot}$ halos, while the lower curves use a flat concentration-mass relationship below $10^{5} \, M_{\odot}$. Dark matter particles with $\sim$$100$ GeV masses consistently exceed the observed gamma ray spectrum when the annihilation cross section is normalized to produce the observed radio background, regardless of the employed annihilation channel. Only lighter dark matter particles can match the intensity of the observed radio flux without violating these constraints.\footnote{We note that under the same assumptions pertaining to magnetic fields and dark matter distributions, our code produces results that are in good agreement with those of Refs.~\cite{Fornengo:2011cn,Fornengo:2011xk}. In particular, for a uniform 10 $\mu$G field and NFW profiles with spatially uniform boost factors, we calculate radio spectra (and quality of fits) which are very similar to those presented in Ref.~\cite{Fornengo:2011cn}.}

In Fig.~\ref{fits}, we show the range of dark matter masses and annihilation cross sections that can produce the observed isotropic radio emission for several annihilation channels, and compare this to the relevant gamma ray constraints. In particular, we show constraints derived from the isotropic gamma ray background (dashed lines)~\cite{fermibg} and from dwarf spheroidal galaxies (dotted lines)~\cite{dwarf}. For the isotropic gamma ray background constraint, we require only that the dark matter contribution does not exceed the reported error bar in any given energy bin, and adopt a flat concentration-mass relationship below $10^5 M_{\odot}$. This leads to a result that is very similar to the ``conservative'' constraints quoted by the Fermi Collaboration~\cite{cosmo}. For the constraints from dwarf galaxies, we have estimated the electron channel constraints based on those from the muon channel, which have a similar spectral shape (the Fermi collaboration does not provide results for the $e^+ e^-$ case). In the case of dark matter which annihilates equally to $e^+ e^-$, $\mu^+ \mu^-$ and $\tau^+ \tau^-$ (democratic leptons), we estimate the dwarf constraint simply by scaling the tau channel constraint by a factor of 1/3. In each case considered, the constraint from dwarf galaxies is more restrictive than that from the isotropic gamma ray background.

From the left frames of Fig.~\ref{fits}, we see that if the Via Lactea substructure model is adopted, gamma ray constraints from dwarf galaxies exclude the vast majority of dark matter parameter space which could potentially explain the observed isotropic radio excess (only a small region near $m_{\rm DM}\sim 5$ GeV annihilating to $e^+ e^-$ survives). If we instead consider the Aquarius-motivated substructure model, more possibilities appear. In particular, annihilations to electrons, muons, or to democratic leptons can each potentially produce the observed radio background without violating any gamma ray constraints.  It is interesting to note that the range of annihilation cross sections required to generate the radio background in the $e^+ e^-$ and democratic leptons cases is very similar to that predicted for a similar thermal relic ($\sigma v\sim 3\times 10^{-26}$ cm$^3$/s).

We emphasize that uncertainties in the magnetic fields and dark matter distributions can shift the annihilation cross sections required to generate the observed radio excess by a factor of a few in either direction. In the next section, we will consider the impact of these uncertainties more closely.

\section{Uncertainties}
\label{uncertainties}

\begin{figure*}[t]
\centering
\includegraphics[angle=0.0,width=3.55in]{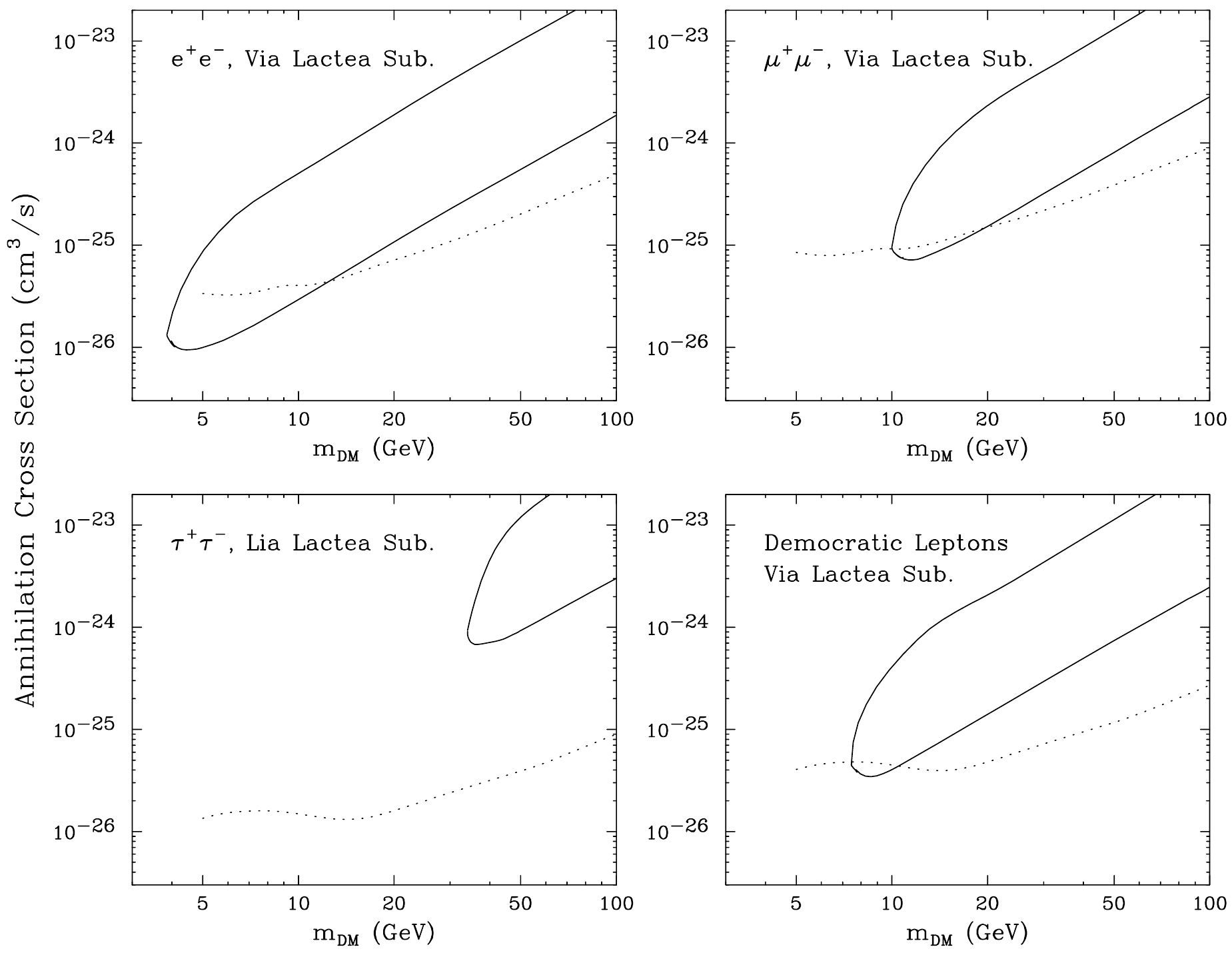}
\hspace{-0.5cm}
\includegraphics[angle=0.0,width=3.55in]{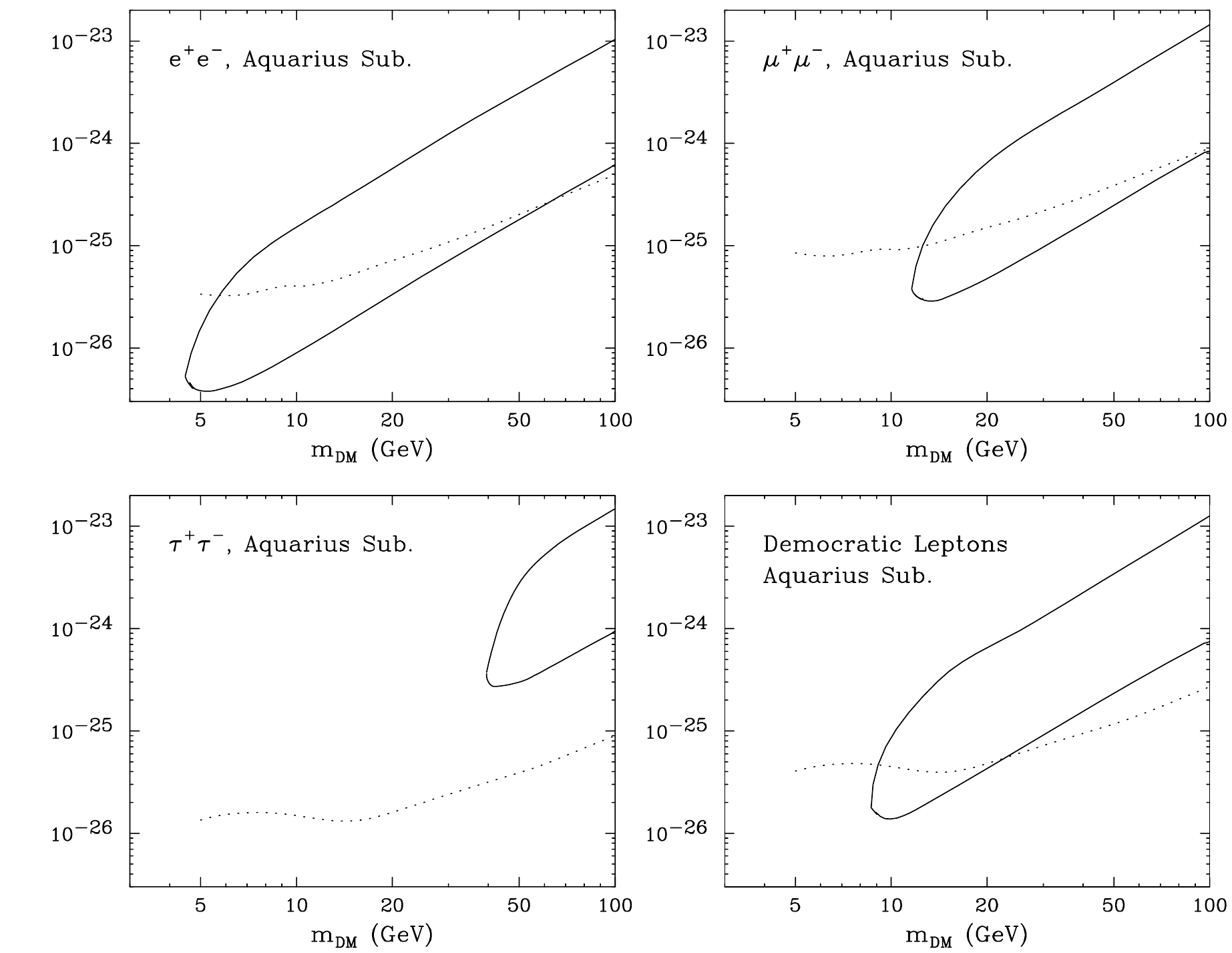}
\caption{As in Fig.~\ref{fits}, but over a range of magnetic field and dark matter distribution models. See Sec.~\ref{uncertainties} for more details.}
\label{fitsunc}
\end{figure*}

In the calculations described in the previous sections of this paper, we adopted specific but reasonable models to describe the distributions of magnetic fields and dark matter. Under those assumptions, we showed that annihilating dark matter particles could potentially account for the observed excess of isotropic radio emission, although only for dark matter models within a relatively narrow range of parameter space. In particular, under those assumptions, the radio background could originate from dark matter annihilations only if the dark matter annihilates mostly to electrons or muons, and with a mass in the range of approximately 4-20 GeV or 14-40 GeV, respectively. In this section, we consider the uncertainties involved in these assumptions and assess whether other reasonable models for the magnetic fields and dark matter distributions might open any other significant regions of dark matter parameter space capable of generating the observed radio background.

In Sec.~\ref{calc}, we described our default magnetic field model for cluster-scale and galaxy-scale halos. In particular, for very massive halos ($M>10^{13} \, M_{\odot}$) we adopted a magnetic field model described by the parameters $B_0=10 \, \mu$G and $\eta=0.5$ (see Eq.~\ref{bcluster}). In this section, we will instead consider the following range for these parameters: $B_0=5-20 \, \mu$G and $\eta=0.5-0.8$. For galaxy-scale halos ($10^{10}-10^{13} \, M_{\odot}$), we consider the range of $B_0=5.9-9.5 \, \mu$G, and include models both with and without a strongly increasing field in the inner region of such halos (see Eq.~\ref{bgalaxy} and text that follows). We also consider halo profiles with inner slopes of $1.0-1.2$ and $1.0-1.4$ for cluster-scale and galaxy-scale halos, respectively.

In Fig.~\ref{fitsunc}, we show how relaxing these assumptions impacts our results. This figure is similar to Fig.~\ref{fits}, but here we have marginalized over the magnetic field models and dark matter distributions described in the previous paragraph. As expected, this leads to broader regions of dark matter parameter space which could potentially account for the observed radio excess. These variations in our underlying assumptions, however, do not qualitatively change our conclusions. Annihilations to $\tau^+ \tau^-$, for example, still cannot provide a good fit to the observed excess without also exceeding gamma-ray constraints (annihilations to quarks or gauge bosons also continue to fail to provide a good fit to the observed radio spectrum). Instead, we find that only dark matter particles which annihilate significantly to electrons and/or muons can potentially account for this signal, although with a slightly larger range of masses than were found in the previous section. The range of annihilation cross sections needed to normalize the synchrotron signal to account for the observed radio excess is also now expanded, covering $\sigma v \approx (0.4-20)\times 10^{-26}$ cm$^3$/s for annihilations to $e^+ e^-$, $\sigma v \approx (3-30)\times 10^{-26}$ cm$^3$/s for annihilations to $\mu^+ \mu^-$, and $\sigma v \approx (1-4.5)\times 10^{-26}$ cm$^3$/s for annihilations to democratic lepton final states (after accounting for gamma ray constraints). Note that in Fig.~\ref{fitsunc}, we do not show the constraints from the isotropic gamma ray background, as these depend on the halo profile shapes which we are now marginalizing over.

\section{Discussion and Conclusions}

A number of observations at frequencies between 20 MHz to 10 GHz~\cite{arcade,roger,maeda,haslam,reich} have revealed the presence of a significant isotropic radio background~\cite{arcadeinterpretation}. While the origin of this emission is currently unknown, it lies in significant excess of expectations from both Galactic sources as well as unresolved extragalactic radio source populations. While a variety of astrophysical sources have been considered (quasars, supernovae, etc.), none appear to be able to account for the observed emission. As previously pointed out by Fornengo {\it et al.}~\cite{Fornengo:2011cn,Fornengo:2011xk}, halos of annihilating dark matter provide a collection of very faint and numerous radio sources, and thus represent an interesting possibility for the origin of the excess isotropic radio emission.

The spectral shape of the observed radio background is quite hard, which limits the range of dark matter models which could potentially account for this emission. In particular, we find that only annihilations to leptonic final states can produce the observed spectrum. Furthermore, dark matter particles which are heavier than $\sim$50 GeV consistently exceed gamma ray constraints if their annihilation cross section is normalized to generate the radio background. After such constraints are taken into account, we find that dark matter particles with a mass of approximately 4-40 GeV annihilating to $e^+ e^-$, 12-50 GeV annihilating to $\mu^+ \mu^-$, or 8-18 GeV annihilating equally to $e^+ e^-$, $\mu^+ \mu^-$ and $\tau^+ \tau^-$ can provide a good fit to the observed isotropic radio excess, depending on the magnetic fields and dark matter distributions that are assumed. For such models, the annihilation cross section required to normalize the synchrotron signal to the observed excess is $\sigma v \approx (0.4-30)\times 10^{-26}$ cm$^3$/s, which is similar to the value predicted for a simple thermal relic ($\sigma v \approx 3\times 10^{-26}$ cm$^3$/s). 

We point out that if dark matter annihilations are responsible for the observed excess radio emission, then a significant fraction of the isotropic gamma ray background observed by Fermi must also result from dark matter. This result is presently consistent with Fermi observations of the anisotropy power spectrum for extragalactic dark matter annihilation, which allow up to 83\% of the extragalactic isotropic background to result from dark matter annihilations (in contrast, the contribution from blazars is constrained to make up no more than 19\% of the diffuse gamma ray emission)~\citep{fermi_anisotropy}.

The characteristics of a dark matter particle required to generate the observed isotropic radio excess are very similar to those required to account for the gamma ray emission observed from the Galactic Center~\cite{gc}, and for the synchrotron emission from the Milky Way's radio filaments~\cite{filaments} and the Inner Galaxy (the ``WMAP Haze'')~\cite{10haze}. In particular, each of these signals can be explained by an approximately 10 GeV dark matter particle which annihilates to electrons, muons, and taus with a total cross section of $\sigma v \approx (1-2)\times 10^{-26}$ cm$^3$/s (assuming a local dark matter density of $0.3\,\,{\rm GeV/cm}^3$). This choice of mass, annihilation channels, and annihilation cross are well matched to those required to generate the excess isotopic radio emission (see the lower right frame of Fig.~\ref{fitsunc}). This range of dark matter masses is also similar to that required to generate the anomalous signals reported by the DAMA/LIBRA~\cite{dama}, CoGeNT~\cite{cogent} and CRESST-II~\cite{cresst} collaborations.

\bigskip
\bigskip
\bigskip

{\it Acknowledgements}:  We would like to thank Doug Finkbeiner and Albert Stebbins for useful discussions. DH is supported by the US Department of Energy. TS is supported by NSF grant PHY-0969448.

\end{document}